\documentclass[11pt, twocolumn]{article}

\usepackage{textcomp}
\usepackage{placeins}
\usepackage{graphicx}
\usepackage{amsmath}
\usepackage[version=4]{mhchem}
\usepackage{siunitx}
\usepackage{longtable,tabularx}
\usepackage{color}

\usepackage{authblk}

\makeatletter
\renewcommand{\maketitle}{\bgroup\setlength{\parindent}{0pt}
\begin{flushleft}
  \textbf{\@title}

  \@author
\end{flushleft}\egroup
}
\makeatother

\title{\Huge Atmospheric Density Model Optimization and Spacecraft Orbit Prediction Improvements Based on Q-Sat Orbit Data  \vspace{0.5cm}}
\author{\Large Zhaokui Wang, Yulin Zhang, Guangwei Wen, Shunchenqiao Bai, Yingkai Cai, Pu Huang, Dapeng Han, Yunhan He \normalsize \hspace{0.5cm}}

\date{\today}

\usepackage[showframe=false, margin=.75in]{geometry}

\usepackage{abstract}
\renewcommand{\abstractname}{}    

\usepackage{lipsum}

\usepackage{afterpage}

\renewenvironment{abstract}
 {\small
  \begin{center}
  \bfseries \abstractname\vspace{-.5em}\vspace{0pt}
  \end{center}
  \list{}{%
    \setlength{\leftmargin}{10mm}
    \setlength{\rightmargin}{\leftmargin}%
  }%
  \item\relax}
 {\endlist}

\usepackage[utf8]{inputenc}

\usepackage{wrapfig}
\usepackage[lofdepth,lotdepth]{subfig}
\usepackage{booktabs}

\usepackage{rotating}

\usepackage{caption}

\usepackage{csquotes}

\usepackage{todonotes}

\usepackage{stfloats}

\usepackage[T1]{fontenc}
\usepackage{textcomp}
\usepackage{times}

\usepackage{framed} 

\usepackage[citestyle=nature,
			bibstyle=nature,
			url=false,
			backend=bibtex,
			doi=true,
                        ]{biblatex}
	\renewbibmacro*{volume+number+eid}{%
  \printfield{volume}%
  \setunit*{\addnbspace}
  \printfield{number}%
  \setunit{\addcomma\space}%
  \printfield{eid}}
\DeclareFieldFormat[article]{number}{\mkbibparens{#1}}
\usepackage{color}
\definecolor{black}{gray}{0} 

\usepackage[colorlinks=true,linkcolor=black,citecolor=black]{hyperref}

\usepackage{tabularx}
\newcolumntype{b}{X}
\newcolumntype{s}{>{\hsize=.5\hsize}X}

\begin{document}

\twocolumn[
\begin{@twocolumnfalse}
\maketitle
\begin{abstract}
Atmospheric drag calculation error greatly reduce the low-earth orbit spacecraft trajectory prediction fidelity.
To solve the issue, the "correction - prediction" strategy is usually employed.
In the method, one parameter is fixed and other parameters are revised by inverting spacecraft orbit data.
However, based on a single spacecraft data, the strategy usually performs poorly as parameters in drag force calculation are coupled with each other, which result in convoluted errors.
A gravity field recovery and atmospheric density detection satellite, Q-Sat, developed by xxxxx Lab at xxx University, is launched on August 6th, 2020.
The satellite is designed to be spherical for a constant drag coefficient regardless of its attitude.
An orbit prediction method for low-earth orbit spacecraft with employment of Q-Sat data is proposed in present paper for decoupling atmospheric density and drag coefficient identification.
For the first step, by using a dynamic approach-based inversion, several empirical atmospheric density models are revised based on Q-Sat orbit data.
Depending on the performs, one of the revised atmospheric density model would be selected for the next step in which the same inversion is employed for drag coefficient identification for a low-earth orbit operating spacecraft whose orbit needs to be predicted.
Finally, orbit prediction is conducted by extrapolation with the dynamic parameters in the previous steps. 
Tests are carried out with the proposed method by using a GOCE satellite 15-day continuous orbit data.
Compared with legacy ``correction - prediction'' method in which only GOCE data is employed, the accuracy of the 24-hour orbit prediction is improved by about 171m the highest for the proposed method.
14-day averaged 24-hour prediction precision is elevated by approximately 70m.
\end{abstract}
\end{@twocolumnfalse}
]

\section{Introducton}
\label{secintro}

Highly accurate low-earth orbit (LEO) operating satellite orbit prediction method are urgently demanded in nowadays as it has a wide range applications such as providing trajectory prediction for optical monitoring and satellite laser ranging \cite{jaggi_goce_2011}, collision possibility analysis as spacecrafts and space debris in LEO increase rapidly \cite{Williamsen2020}.
Most of the LEO spacecraft orbit prediction errors result from inaccurate drag calculation especially atmospheric drag.
It is common to use "correction-prediction" strategy to elevate prediction accuracy.
For the step of "correction" , spacecraft orbital data is employed for inversion to correct parameters in the drag model such as atmosphere density, drag coefficient, and gas-surface interaction parameters.
Usually, one parameter is fixed for correction of other parameters.
In the step of "prediction", numerical integration is performed to obtain orbit extrapolation and the corrected parameters in previous step are applied \cite{Montenbruck2000}.
However, coupling of the parameters in the atmospheric drag model results in convoluted uncertainties for the "correction-prediction" strategy \cite{Thayer_2021_Remaining}.
The corrected parameter "absorbs" the uncertainties in other parameters.
Generally, for spacecraft orbits at altitude of 400km-500km in the next 24 hours, the accuracy of the forecast using "correction-prediction" scheme can be reached between 30m to 60m. 
For the spacecraft at altitude of 300km-400km, the prediction accuracy is around 200m to 300m \cite{Wei2018Affection,Wang2018Accuracy}. 
For lower altitudes, (less than 300km) trajectory prediction accuracy drops to 1000m to 1500m \cite{jaggi_goce_2011}. 
It can be observed that as the rarefied gas density increases thus atmospheric drag ascends, the efficacy of the "correction-forecast" strategy decreased.
To unravel the interdependency, investigations were conducted to evaluate spacecraft drag coefficient individually.
Nevertheless, factors such as temperature, molecular composition, degree of particles adsorption on the surface, surface material cause a virational drag coefficient \cite{ray_drag_2020,crisp_orbit_2021}.
It is reported that because of drag coefficient erroneous, atmospheric models there were developed by satellite orbital behavior overestimate or underestimate rarefied gas density tremendously \cite{moe_gassurface_2005}.
To have a better estimation for the drag coefficient, diffuse reflection with incomplete accommodation \cite{bernstein_evidence_2020}, Cercignani-Lampis-Lord \cite{mehta_comparing_2014} and other scattering models are usually employed \cite{mehta_modeling_2014}. 
Furthermore, selection of empirical atmosphere density model (EADM) are vital in  drag calculation and therefore the spacecraft orbit forecast \cite{Wei2018Affection,Qiu2006Comparison}.
At present, EMADs that are commonly employed are, for example, Jacchia models, NRLMSISE-00, JB2008, DTM2000, etc. 
Each model perform differently and no model could cover all the aspects (solar flux, geomagnetic index, atmospheric winds, distribution of molecular concentrations, etc.) in space environment \cite{Jin_2020_Atmospheric}.
For instance, NRLMSISE-00 is less sensitive to the parameter such as solar activity, geomagnetic activity, latitude, and longitude than that of the actual atmosphere \cite{Miao2018}. 
Jacchia models outperform the MSIS models when considering the impact of solar activity for orbital prediction \cite{healy_effects_2004}. 
For altitude less than 500km, JB2008 and Jacchia71 have a good performance in tasks like density representation and orbit prediction \cite{Wei2018Affection,Zhao2019PHDTHESIS}.
Revising a EADM is an important mean to improve orbit prediction accuracy.
Inversion methods based on semi-major axis attenuation or non-conservative forces \cite{Huang2018Method,Li2017,Kuang2014} are often performed by using {\em a priori} orbit data for atmosphere density recovery. 
However, such methods cannot provide atmosphere density forecasts \cite{gondelach_atmospheric_2020}. 
Based on the dynamic approach-based inversion method, Wang proposed an technique for atmospheric density forecast \cite{Wang2019}. 
The method was later improved by Zhang et al. \cite{Zhang2018-Atmospheric} as thermal parameters are involved in.
Recently, efforts has been devoted for diminishing the convolution effect as can be seen work by Ray et al. \cite{Ray_2020Drag,Ray_2021Inverting} who tried to estimating atmospheric density and drag coefficient simultaneously by spacecraft tracking data.
Fourier series was employed for filtering uncertainties in the drag coefficient.

\begin{figure}
\centerline{
\includegraphics[width=3.in]{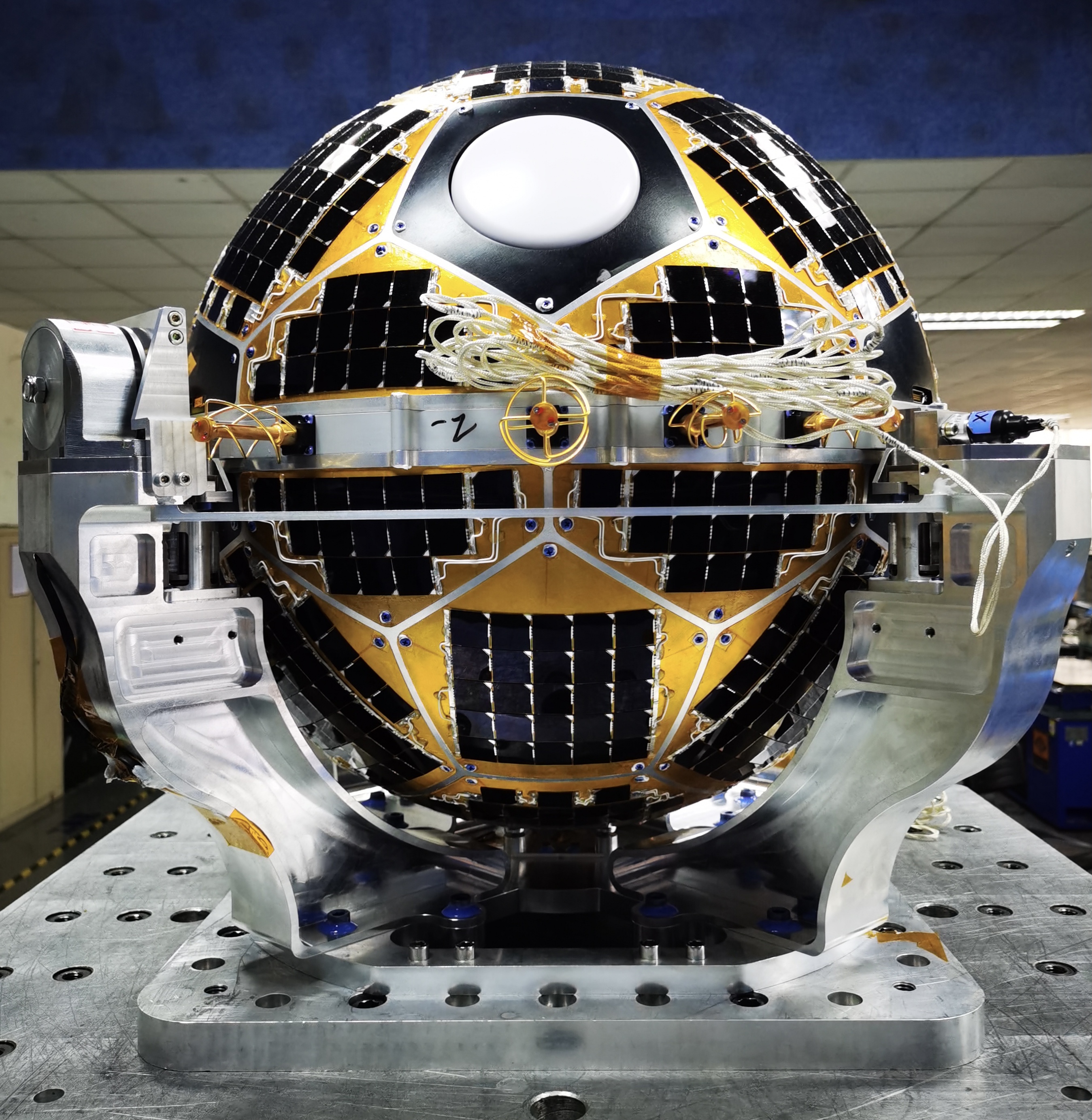}\noindent
}
\caption{
Q-Sat with seperation system.
}
\label{Q-Sat}
\end{figure}

The Q-sat satellite, also known as Gravity and Atmospheric Science Satellite of xxx University, is a micro spherical satellite developed by the xxxxx Lab (xxxxx) at xxx University \cite{zhao_spherical_2019}.
The satellite has a net weight of 21.2kg, a diameter of about 510mm.
It operates in a sun-synchronous orbit at initial altitude of approximately 500km. 
Since it designed to be nearly spherical, the Q-sat can be considered to have a constant drag coefficient of $2.2$ regardless its orientation based on molecular dynamics numerical simulation and experiments \cite{Wang2019,Zhao2019PHDTHESIS}.
Dual frequency, carrier phase differential global navigation satellite system (GNSS) receiver is equipped and both Global Positioning System (GPS) and Beidou Navigation Satellite System (BDS) signal can be used for positioning.
After post-processing, the orbit data could have accuracy of centimeter level.
The xxxxx lab fetches the Q-Sat orbit data on a daily basis.
Image of the Q-Sat with separation system is provided in Fig. \ref{Q-Sat}.
The separation system is intended to mount on the launch vehicle \cite{HE2021259,YUNHAN2021180}.
More detail of the satellite can be found in Ref. \cite{zhao_spherical_2019,HE2021259}.

In current paper, an orbit prediction method which amis to improve prediction accuracy is proposed. 
In the first step, multiple EADMs would be modified by using Q-sat orbit data and a dynamic approach-based inversion.
One of the EADM that performs the best would be selected.
For demonstration of the proposed method, the Jacchia-Roberts atmospheric density model are modified in current paper.
Next, by using the same inversion method, the drag coefficient for the LEO spacecraft which orbit needs to be forecast is identified.
In this way, transmission error caused by using the same satellite orbit data can be reduced and the updated drag coefficient grounds an improved trajectory prediction accuracy. 
Tests are carried out by employing a Gravity Field and Steady-State Ocean Circulation Explorer (GOCE) satellite 15-day continuous orbit data.
Obit prediction for GOCE satellite was conducted by proposed dynamic approach-based orbit prediction procedure.
Legacy "correction-prediction" strategy was also employed for comparison.
Discussions and conclusion are provided in the end.

\section{Methodology}

\subsection{Q-Sat Trajectory Based LEO Spacecraft Orbit Prediction Procedure}

The proposed LEO spacecraft orbit prediction method can be outlined in 4 steps as shown in the following.
\begin{enumerate}
\item Modification of EADMs: 
A dynamic approach-based inversion method is first performed by using Q-sat orbit data for EADM revisions.
On-board dual-frequency, carrier phase differential GNSS receiver makes orbit determination centimeter level precision possible.
Several EADMs would be modified in current step.
\item Selection of modified EADM: 
a optimized atmospheric density model that performs the best is selected.
\item Identification of drag coefficient: 
by involving the revised EADM and orbit data of the spacecraft whose trajectory needs to be predicted, the same dynamic approach-based inversion method employed on step 1 is performed again to obtain a drag coefficient for the spacecraft.
The spacecraft orbit data for drag coefficient identification and the Q-Sat orbit data for EADM revision should from the same day.  
\item Orbit forecast: based on the revised EADM and drag coefficient obtained in previous steps, orbit prediction is achieved by extrapolation.
\end{enumerate}
It should be noted that because of the space environment variation, the obtained EADM and drag coefficient is considered to be valid for a short period of time.
Inversion for dynamic parameters should be carried out again if it is for orbit prediction long time after.

\subsection{Inversion for EADM Optimization and Drag Coefficient Identification}

A dynamic approached-based inversion would be introduced in this section.
It can be employed for inverting different parameters individually by using spacecraft orbital data.
Acceleration generated by various forces on a LEO spacecraft which orbits around the earth can be expressed as,
\begin{equation}
\ddot{{\bf r}}=-\frac{G M}{r^{3}} {\bf r}+f(t, {\bf r}, \dot{{\bf r}}, {\bf p}),
\label{acceleration}
\end{equation}
where ${\bf r}$ is the satellite position vector, $\dot{{\bf r}}$ is satellite velocity vector, $\ddot{{\bf r}}$ is total acceleration vector for the spacecraft at epoch $t$, $p$ is the dynamic parameter vector, and $GM$ is the earth gravitational constant. 
The first term on the right is two-body gravitational acceleration, and the second term is the acceleration produced by the perturbation.
The actual spacecraft motion state at epoch $t$ can be represented as,
\begin{equation}
{\bf X}(t)= \left ( {\bf r}^{\mathrm{T}}(t) ,\dot{{\bf r}}^{\mathrm{T}}(t) \right )^{\mathrm{T}},
\end{equation}
and can be obtained from spacecraft orbit data.
Spacecraft motion state obtained by a dynamic approach-based integration according to Eq. \ref{acceleration}  can be presented as,
\begin{equation}
\tilde{{\bf X}}(t)= \left ( \tilde{ {\bf r}}^{\mathrm{T}}(t) ,\tilde{\dot{{\bf r}}}^{\mathrm{T}}(t) \right )^{\mathrm{T}}.
\end{equation}
Motion state calculated by the integration is function of the initial motion state  and dynamic parameters,
\begin{equation}
\tilde{{\bf X}}(t)=\tilde{{\bf X}}({\bf r}_{o},\dot{{\bf r}}_{o},{\bf p},t).
\end{equation}
Next, difference between satellite actual motion state and motion state obtained by the integration at epoch $t_n$ can be linearized as,
\begin{equation}
\begin{aligned}
\Delta {\bf X}_{t_n} &={\bf X}(t_n)-\tilde{{\bf X}}\left({\bf X}_{o}, {\bf p}, t_n\right) \\
&\left.\approx \frac{\partial {\bf X}(t_n)}{\partial {\bf X}_o}\right|_{{\bf X}_o}\left(\tilde{{\bf X}}_o-{\bf X}_o\right)+\left.\frac{\partial {\bf X}(t_n)}{\partial {\bf p}}\right|_{{\bf p}}(\tilde{{\bf p}}-{\bf p}) \\
&=\left.\frac{\partial {\bf X}(t_n)}{\partial {\bf X}_o}\right|_{{\bf X}_o} \Delta {\bf X}_{o}+\left.\frac{\partial {\bf X}(t_n)}{\partial {\bf p}}\right|_{{\bf p}} \Delta {\bf p},
\end{aligned}
\end{equation}
where $\partial {\bf X}(t_n) / \partial {\bf X}_o $ is transtion matrix, it can be written by ${\bf \varPhi}_{t_n}$ and the matrix represents how the initial motion state error affect the subsequent motion states, $\partial {\bf X}(t_n) / \partial {\bf p}$ is sensitivity matrix and it can be aliased by ${\bf S}_{t_n}$. 
The sensitivity matrix shows the impact of dynamic parameter errors on the subsequent motion states \cite{Montenbruck2000}.
Finally, corrections for initial state and dynamic parameters can be obtained by the least square method \cite{Zhao2019PHDTHESIS} since it is an over-determined system,
\begin{equation}
\left[\begin{array}{c}
\Delta {\bf X}_{0} \\
\Delta {\bf p}
\end{array}\right]=\left[({\bf \varPhi}, {\bf S})^{\mathrm{T}}({\bf \varPhi}, {\bf S})\right]^{-1}({\bf \varPhi}, {\bf S})^{\mathrm{T}} \Delta {\bf X},
\label{LSF}
\end{equation}
where,
\begin{equation}
\Delta {\bf X} = \left[\begin{array}{c}
\Delta {\bf X}_{t_1} \\
\Delta {\bf X}_{t_2} \\
\vdots\\
\Delta {\bf X}_{t_n} 
\end{array}\right],
\left ( {\bf \varPhi},{\bf S} \right)= \left[\begin{array}{c}
\left ( {\bf \varPhi}_{t_1},{\bf S}_{t_1} \right) \\
\left ( {\bf \varPhi}_{t_2},{\bf S}_{t_2} \right) \\
\vdots\\
\left ( {\bf \varPhi}_{t_n},{\bf S}_{t_n} \right)
\end{array}\right].
\end{equation}

EADM can be used to describe earth atmosphere status and how it changes \cite{Zhang2010Introduction}, generally it is a function of temperature, atmospheric particle composition, solar activity, geomagnetic activity and other variables. 
For current investigation, attentions should be paid on parameter sensitivity thus the orbit prediction accuracy.
In present paper, in order to present a EADMs modification process, Jacchia-Roberts model is chosen. 
It should be noted that in actual spacecraft orbit prediction improvement workflow, several EADMs should be modified and compared in terms of orbit prediction performance.

For temperature at height, $h$, above 125km,
\begin{equation}
\begin{aligned}
&T_h=T_{\infty}\\
&-(T_{\infty}-T_{x})\exp \left (-(\frac{T_x-183}{T_{\infty}-T_x})(\frac{h-125}{35})(\frac{L}{R_a+h}) \right),
\end{aligned}
\label{temperatureProfile}
\end{equation}
where $T_{\infty}$ is exospheric temperature, $T_x$ is the temperature at 125km, $R_a$ is the polar radius of the earth and it is defined as $R_a=6356.766$km, and $L$ is a corrective factor.
Detailed descriptions can be obtain in paper by Roberts \cite{roberts_analytic_1971}.
For current EADM modification, $L$ is expressed as a polynomial function of $T_{\infty}$ \cite{LiJisheng1995},
\begin{equation}
L=\sum\limits_{i=i}^5 l_{i} T_{\infty}^{i-1},
\end{equation}
in which $l_i$ is the dynamic model parameters and ${\bf l}=[l_1,l_2,\cdot\cdot\cdot, l_5]$.
Therefore, the sensitivity matrix for the acceleration by atmospheric drag is,
\begin{equation}
{\bf S}(t)=\frac{\partial {\bf a}_{ato}}{\partial {\bf l}}=-\frac{1}{2} C_d \frac{A}{m} \frac{\partial \rho}{\partial L} \frac{\partial L}{\partial {\bf l}} v_r {\bf v}_r,
\label{sensitivityMatrix}
\end{equation}
where $C_d$ is the spacecraft drag coefficient, $A$ is the cross-sectional area, $m$ is the spacecraft mass, ${\bf v}_r$ is spacecraft relative velocity.
In Jacchia-Roberts model, atmospheric density can be obtained as \cite{Montenbruck2000,roberts_analytic_1971,LiJisheng1995},
\begin{equation}
\rho(h)=\rho_s \Delta \rho + \Delta \rho_{He},
\end{equation}
where $\rho_s$ is the calculated standard atmospheric density, $\Delta \rho$ is the corrected density under the effects of geomagnetism, half-year period, seasonal latitude, etc., and $\Delta \rho_{He}$ is the corrected density for helium.
Next, partial derivative respect to corrective factor $L$ is performed as,
\begin{equation}
\frac{\partial \rho}{\partial L}=\frac{\partial \rho_s}{\partial L}\Delta \rho +\frac{\partial \rho_{He}}{\partial L}.
\end{equation}

The term of $\partial \rho_s / \partial L$ can be expanded as,
\begin{equation}
\begin{aligned}
\frac{\partial \rho_{\mathrm{s}}}{\partial L}=&\sum_{i=1}^{5} \rho_{i}(125)[ \frac{\partial\left(\frac{T_{x}}{T_{h}}\right)^{1+\alpha_{i}+\gamma_{i}}}{\partial L}\left(\frac{T_{\infty}-T_{h}}{T_{\infty}-T_{x}}\right)^{\gamma_{i}}\\
&+\left(\frac{T_{x}}{T_{h}}\right)^{1+\alpha_{i}+\gamma_{i}} \frac{\partial\left(\frac{T_{\infty}-T_{h}}{T_{\infty}-T_{\mathrm{x}}}\right)^{\gamma_{i}}}{\partial L} ]+\frac{\partial \rho_{6}(h)}{\partial L},
\end{aligned}
\label{partailToL}
\end{equation}
where $i=1,2,\cdot\cdot\cdot,5$ and it represent $N_2$, $Ar$, $He$, $O_2$, and $O$ correspondingly, $\rho_6$ is the density for hydrogen particles,
\begin{equation}
\gamma_i=\frac{M_i g_o R_a^2}{ R L T_{\infty}} \left( \frac{T_{\infty}-T_x}{T_x-183}\right)  \left( \frac{35}{R_a+125} \right),
\end{equation}

\begin{equation}
\begin{aligned}
&\frac{\partial\left(\frac{T_{x}}{T_{h}}\right)^{1+\alpha_{i}+\gamma_{i}}}{\partial L}=\left(\frac{T_{x}}{T_{h}}\right)^{1+\alpha_{i}+\gamma_{i}}\\
& \left[\frac{\partial \gamma_{i}}{\partial L} \ln \left(\frac{T_{x}}{T_{h}}\right)+\left(1+\alpha_{i}+\gamma_{i}\right) \frac{T_{h}}{T_{x}} \frac{\partial\left(\frac{T_{x}}{T_{h}}\right)}{\partial L} \right],
\end{aligned}
\end{equation}

\begin{equation}
\begin{aligned}
\frac{\partial \gamma_{i}}{\partial L}=-\frac{M_{i} g_{0} R_{a}^{2}}{R T_{\infty}} \frac{1}{L^{2}} \left(\frac{T_{\infty}-T_{x}}{T_{x}-183}\right) \left( \frac{35}{R_a+125} \right),
\end{aligned}
\end{equation}

\begin{equation}
\begin{aligned}
&\frac{\partial\left(\frac{T_{x}}{T_{h}}\right)}{\partial L}=\\
&\frac{T_{x}}{T_{h}^{2}}\left(T_{\infty}-T_{x}\right)\\
&\exp \left(-\frac{T_{\infty}-183}{T_{\infty}-T_{x}} \frac{h-125}{35} \frac{L}{R_a+h}\right)\\
&\left(-\frac{T_{\infty}-183}{T_{\infty}-T_{x}} \frac{h-125}{35} \frac{1}{R_a+h}\right),
\end{aligned}
\end{equation}

\begin{equation}
\begin{aligned}
&\frac{\partial \ln \frac{T_{\infty}-T_{h}}{T_{\infty}-T_{x}}}{\partial L}=\\
&\frac{T_{\infty}-183}{T_{h}-T_{\infty}} \\
&\exp \left(-\frac{T_{\infty}-183}{T_{\infty}-T_{x}}\frac{h-125}{35} \frac{L}{R_a+h}\right)\\
&\left(T_{\infty}-183\right) \frac{h-125}{35} \frac{1}{R_a+h},
\end{aligned}
\end{equation}
$\alpha_i$ is the diffusion coefficient, $M_i$ is molar mass of the gas component $i$, $g_o$ is the gravitational acceleration at sea level, and $R$ is the gas constant, and finally for the last term in Eq. \ref{partailToL} which considers the hydrogen particles,
\begin{equation}
\begin{aligned}
\frac{\partial \rho_{6}}{\partial l}=&\rho_{6}(500) [\frac{\partial\left(\frac{T_{\infty}-T_{s}}{T_{h}}\right)^{1+\alpha_{6}+\gamma_{6}}}{\partial l}\left(\frac{T_{\infty}-T_{h}}{T_{s}}\right)^{\gamma_{6}}\\
&+\left(\frac{T_{\infty}-T_{s}}{T_{h}}\right)^{1+\alpha_{6}+\gamma_{6}} \frac{\partial\left(\frac{T_{\infty}-T_{h}}{T_{s}}\right)^{\gamma_{6}}}{\partial l}]
\end{aligned}
\end{equation}
where $T_s$ is the temperature at 500km according to Eq. \ref{temperatureProfile} and,
\begin{equation}
\begin{aligned}
&\frac{\partial T_s}{\partial L}=\\
&(T_{\infty}-T_x)\exp\left( (\frac{ T_x-183}{T_{\infty}-T_x})(\frac{375}{35}))(\frac{1}{R_a+h})\right),
\end{aligned}
\end{equation}
\begin{equation}
\begin{aligned}
&\frac{\partial T_{h}}{\partial l}=\\
&\left[(T_x-183) (\frac{h-125}{35}) (\frac{1}{R_a+h})\right]\\
&\exp \left(-\frac{T_{\infty}-183}{T_{\infty}-T_{x}} \frac{h-125}{35} \frac{l}{R_a+h}\right),
\end{aligned}
\end{equation}
\begin{equation}
\frac{\partial\left(\frac{T_{\infty}-T_{s}}{T_{h}}\right)}{\partial l}=\frac{-\frac{\partial T_{s}}{\partial l} T_{h}-\left(T_{\infty}-T_{s}\right) \frac{\partial T_{h}}{\partial l}}{T_{h}^{2}},
\end{equation}
\begin{equation}
\begin{aligned}
\frac{\partial\left(\frac{T_{\infty}-T_{h}}{T_{s}}\right)^{\gamma_{6}}}{\partial l}=&\left(\frac{T_{\infty}-T_{h}}{T_{s}}\right)^{\gamma_{6}}\\
&\left[\frac{\partial \gamma_{6}}{\partial l} \ln \frac{T_{\infty}-T_{h}}{T_{s}}+\gamma_{6} \frac{\partial \ln \frac{T_{\infty}-T_{h}}{T_{s}}}{\partial l}\right],
\end{aligned}
\end{equation}
\begin{equation}
\frac{\partial \ln \frac{T_{\infty}-T_{h}}{T_{s}}}{\partial l}=\frac{T_{s}}{T_{\infty}-T_{h}} \frac{-\frac{\partial T_{h}}{\partial l} T_{\mathrm{s}}-\left(T_{\infty}-T_{h}\right) \frac{\partial T_{s}}{\partial l}}{T_{s}^{2}}.
\end{equation}
For the term of $\partial \rho_{He} / \partial L$ which is for Helium density correction,
\begin{equation}
\frac{\partial \Delta \rho_{He} }{\partial L}= \frac{\partial \Delta \rho_{He}(h) }{\partial L} \left[  10^{\Delta log_{10} n_{He}(h)}-1\right].
\end{equation}
Lastly, by substituting the equations above and,
\begin{equation}
\frac{\partial L }{\partial {\bf l}}= [1, T_\infty, T_\infty^2, T_\infty^3, T_\infty^4],
\end{equation}
in Eq. \ref{sensitivityMatrix}, the sensitivity matrix, ${\bf S}(t)$, for atmospheric drag acceleration is obtained.

The input for the proposed dynamic approach-based inversion is a spacecraft orbital data.
In order to optimize a EADM, drag coefficient is fixed to a certain value and corrective parameters for the EADM would be on the output.
The over-determined system is solved by least square fitting as shown in Eq. \ref{LSF}. 
For demonstration of EADM correction in current paper, corrective factors $l_i, i=1,2,\cdot\cdot\cdot,5$ for Jacchia-Roberts are on the output of the inversion.
The practice is especially considered as appropriate for orbit data from spherical satellite, for example the Q-Sat launched by xxxxx at xxx University as its drag coefficient is not relative to its inclination (angle of attack).
After EAMDs are revised and the optimal EAMD is selected, the tracking data of spacecraft which trajectory needs to be predicted is employed as the input of the inversion and revised drag coefficient for the spacecraft will be on the output.
Lastly, the optimized EADM and drag coefficient are involved for orbit extrapolation.

\section{Results}
\subsection{EADM Optimization}
\begin{figure*}
\centerline{
\includegraphics[width=6.5in]{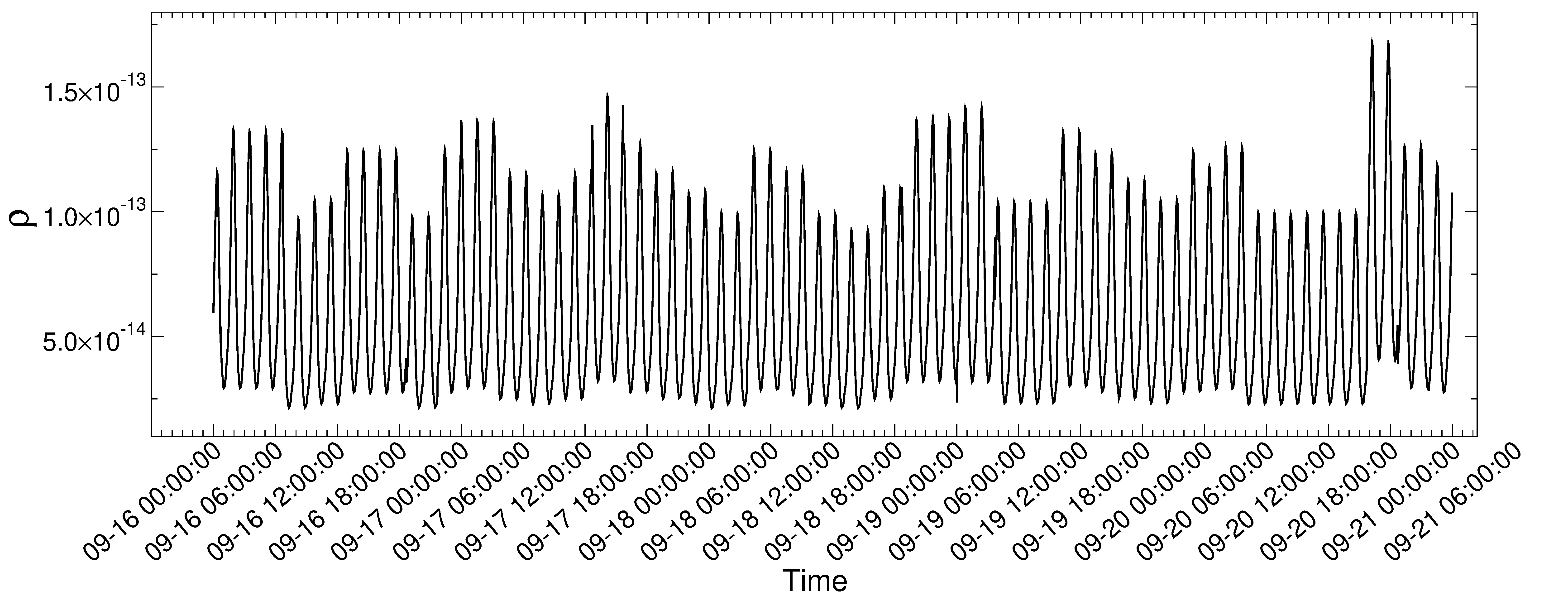}\noindent
}
\caption{
Atmospheric density along Q-Sat orbit by optimized Jacchia-Roberts model.
}
\label{AD}
\end{figure*}

For demonstration for EADM optimization, the Q-Sat orbit data is employed and data for every 24-hour is considered as a data set.
A consecutive 5 days data that expands from September 16 to September 20, 2020 is used. 
After post-processing the orbit data from Q-Sat on-board dual frequency, carrier phase differential GPS/BDS receiver, the obtained orbit data could have centimeter-level accuracy.
In present paper, corrective factors, $l_i, i=1,2,\cdot\cdot\cdot,5$, for Jacchia-Roberts atmospheric model is optimized by the proposed dynamic approach-based inversion method.
Five set of corrective factors are obtained as shown in Tab.\ \ref{CF-EADM}
\begin{table*}
\caption{Corrective factors for Jacchia-Roberts model\label{CF-EADM}}
\begin{tabular}{@{}cccccc@{}}
\hline
Date & \multicolumn{5}{c}{${\bf l}$} \\
\hline
09-16 & $4.8144\times10^3$ & $0.2341$ & $1.5792\times10^{-3}$ & $-1.2525\times10^{-6}$ & $2.4627\times10^{-10}$\\ 
09-17 & $5.4073\times10^3$ & $0.2341$ & $1.5792\times10^{-3}$ & $-1.2525\times10^{-6}$ & $2.4627\times10^{-10}$\\
09-18 & $4.5516\times10^3$ & $0.2341$ & $1.5792\times10^{-3}$ & $-1.2525\times10^{-6}$ & $2.4627\times10^{-10}$\\
09-19 & $5.2256\times10^3$ & $0.2341$ & $1.5792\times10^{-3}$ & $-1.2525\times10^{-6}$ & $2.4627\times10^{-10}$\\
09-20 & $4.7134\times10^3$ & $0.2341$ & $1.5792\times10^{-3}$ & $-1.2525\times10^{-6}$ & $2.4627\times10^{-10}$\\
\hline
\end{tabular}
\end{table*}
By applying the corrective factors, atmospheric density along the Q-Sat orbit is calculated and plotted in Fig. \ref{AD}.
At 16:14:20 (hh:mm:ss) on September 20th, the density is the highest at $1.6807\times10^{-13}kg/m^3$ during the 5 days. 
At the time, the Q-Sat height is about 495.6604km, the geomagnetic index is $K_p \approx 2.3nT$, and the daily average value of solar radiation index is $F_{10.7} \approx 71\times10^{-22}W/(m^2 \cdot Hz)$, which is the largest in 5 days. 
At 00:17:50 on September 18, 2020, the density is $2.0842\times10^{-14} kg/m^3$ which is the smallest in the 5 days.
At the time, the Q-Sat height is about 495.6721km, the geomagnetic index is approximately $0nT$, which is the smallest in 5 days, and the daily average value of solar radiation index is about $70\times10^{-22}W/(m^2 \cdot Hz)$, only slightly higher than $69\times10^{-22}W/(m^2 \cdot Hz)$ on Sept 16th.

In order to examine the proposed inversion for EADM optimization, an arclength of 27 hours orbit was employed and atmospheric densities calculated by optimized EADM are compared.
For the first 27 hours, orbit data from the first day was added with the first 3 hours data on the second day for a total of 27 hours.
For the second 27-hour, orbit data from the last 3 hours of the first day was added at the beginning of the second day.
Next, the 27-hour data is employed for EADM optimization.
The atmospheric densities for the last 3 hours of the first day calculated by EADMs that are optimized by 27-hour data and by 24-hour data were compared.
The same comparisons were done for the atmospheric densities for the first 3 hours of the second day.
As a result, there are in total of eight 3-hour rarefied gas density intervals for comparison from September 16 to September 20 and maximum relative errors and mean relative errors over the 3-hour are presented in Tab. \ref{ADcompare}.
\begin{table*}
\caption{Atmospheric density relative error\label{ADcompare}}
\begin{tabular}{@{}ccc@{}}
\hline
Time &  Maximum relative error(\%) & Mean relative error(\%)\\
\hline
09-16, 21:00-12:00 & 5.2310 & 4.5173\\
09-17, 00:00-03:00 & 4.9710 & 4.3196\\
09-17, 21:00-12:00 & 15.1412 & 13.0713\\
09-18, 00:00-03:00 & 17.8428 & 15.0756\\
09-18, 21:00-12:00 & 12.2994 & 11.0099\\
09-19, 00:00-03:00 & 10.9523 & 9.9109 \\
09-16, 21:00-12:00 & 6.545874 & 5.744469\\
09-17, 00:00-03:00 & 7.0043 & 6.0983\\
\hline
\end{tabular}
\end{table*}
As one can observe in the table, in the 8 comparison intervals, the maximum relative error is a little less than $18\%$, the smallest is around $5\%$. 
For the mean relative errors, the largest is around $15\%$ at the smallest is around $4.3\%$.
The total average of the mean relative errors is about $8\%$. 

\FloatBarrier

\subsection{Drag Coefficient Identification}

For current step, the input for the dynamic approach-based inversion is the orbit data of the LEO spacecraft which orbit prediction is on demand, the drag coefficient for the spacecraft is on the output.
Revised EADM carried out by Q-Sat orbit behavior is employed in the inversion for LEO spacecraft drag coefficient identification.
Not to mention that the revised EADM shall only be used for identifying drag coefficient at the same day.
The revised EADM and redefined spacecraft drag coefficient are produced in a daily basis in order to include variations and anomalies in an entire day and would not be accurate for other days with different space environment.
This practice lead to a potentially better prediction accuracy.
Since it is hard to reach orbital information of a LEO spacecraft that is currently orbiting the earth, for demonstration of the proposed orbit prediction improvement method, the GOCE satellite orbit data is employed.
The satellite data is open to the public and can be downloaded from the European Space Agency website.
The on-board dual frequency GPS receiver provided position data every 10 seconds and the data could have centimeter level accuracy with post-processing. 

\subsubsection{Space Environment Parameters}
\begin{figure}[hhh!]
\centerline{
\includegraphics[width=3.5in]{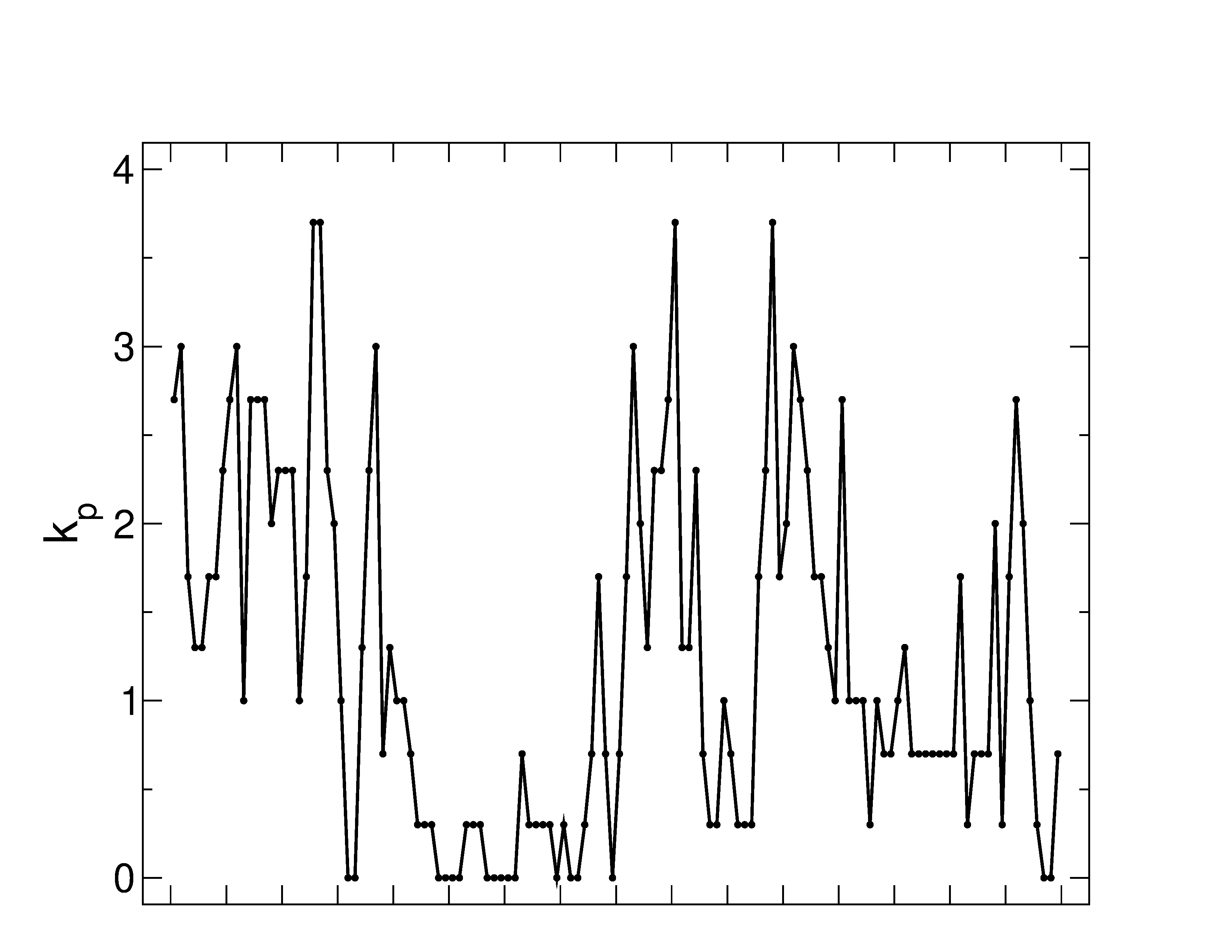}\noindent
}
\centerline{
\includegraphics[width=3.5in]{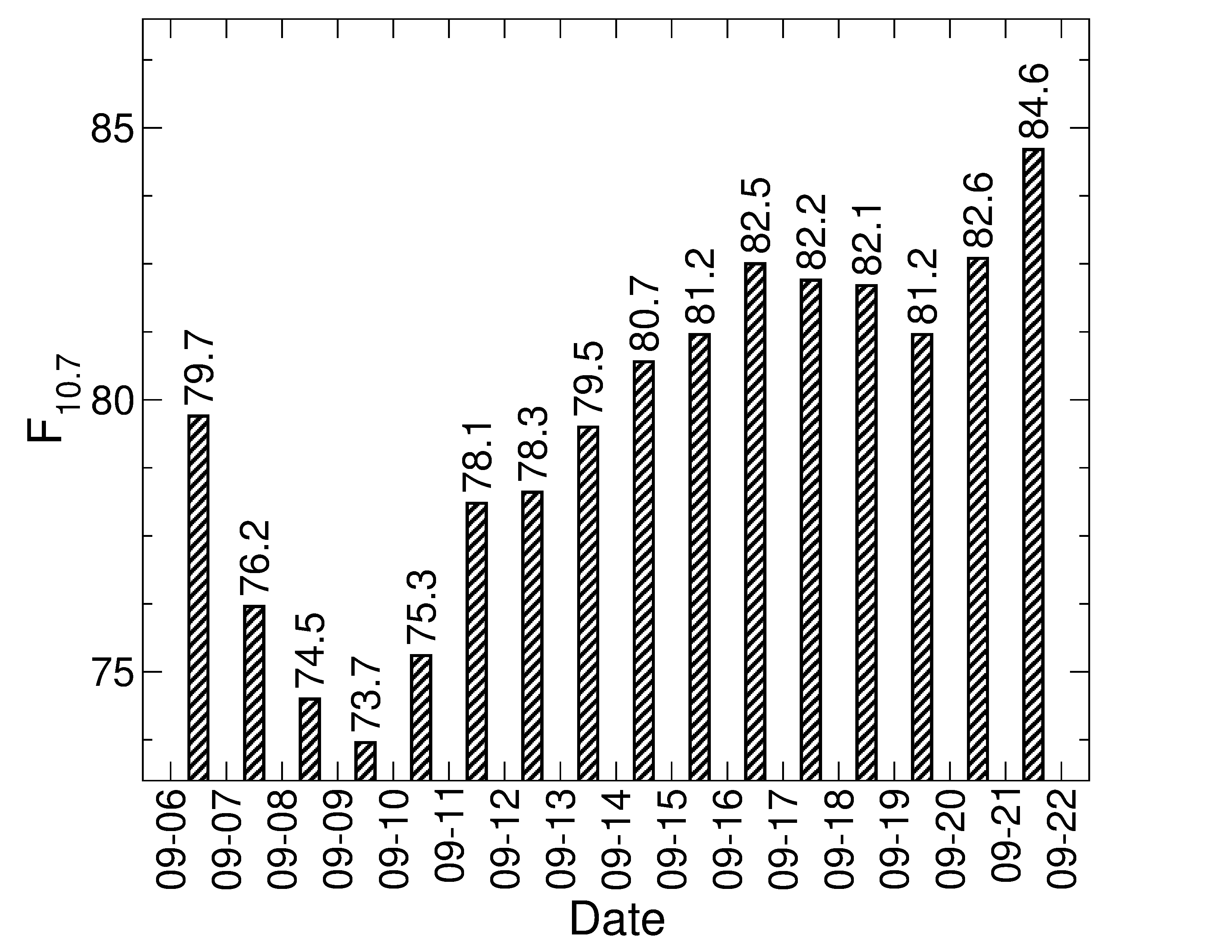}\noindent
}
\caption{
Geomagnetic index, $K_p$, and solar radiation index, $F_{10.7}$, in September of 2010.
}
\label{EPs}
\end{figure}

\begin{figure}[hhh!]
\centerline{
\includegraphics[width=3.5in]{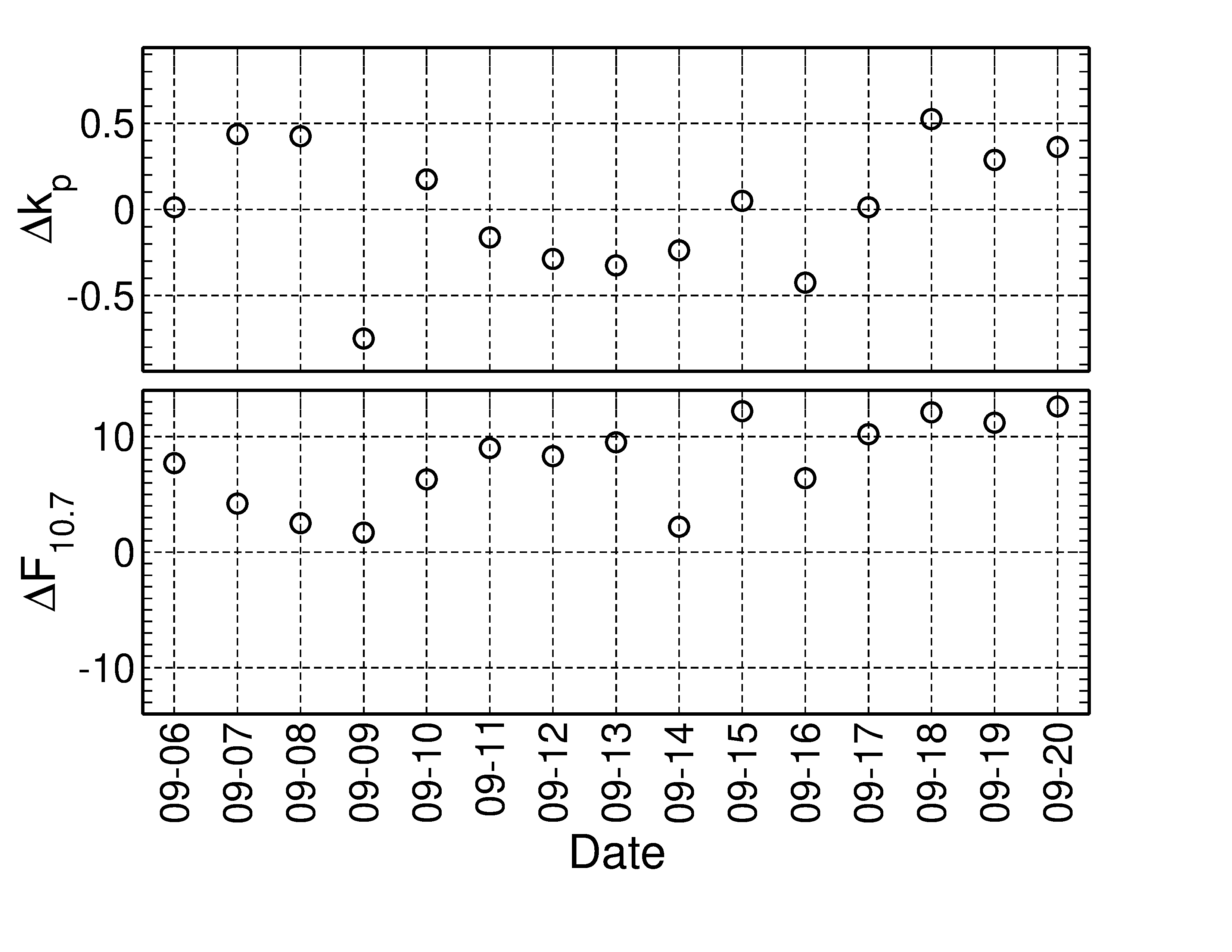}\noindent
}
\caption{
Differences of daily averaged geomagnetic index and solar radiation index between GOCE arc and Q-Sat arc.
}
\label{EP_difference}
\end{figure}

A 15-day continuous orbit for GOCE satellite is used.
The arc extents from September 6th to September 21st in 2010 in which the on-board ion propulsion engine was not active for drag-free flight and the satellite was descending.
The arc of first day is utilized for drag coefficient identification and the arc for the second day is employed for examining the fidelity of the orbit prediction.
Drag coefficient and space environment parameters obtained in the previous day are used in orbit propagation for the current day as the space situation of the current day often show a good match with those of the previous day \cite{jaggi_goce_2011}.
Since the Q-Sat was launched lately in 2020 and the orbit data is employed for EADM optimization, to implement drag coefficient identification for GOCE satellite with orbit data in 2010, the space environment for Q-Sat orbit data this is used for EADM optimization should be analogous to that of GOCE \cite{vallado_critical_2014}.
For the Jacchia-Roberts atmospheric model that is modified in previous section, the geomagnetic index, $K_p$, and the solar radiation index, $F_{10.7}$, are the variables for calculating the exospheric temperature, $T_{\infty}$. 
The GOCE arc for drag coefficient identification should have similar values of $k_p$ and $F_{10.7}$ with the Q-Sat arc which is employed EADM optimization.
If it is for prediction of a spacecraft that is currently orbiting, this step is unnecessary and it is only needed for present demonstration. 
Since usually it is hard to obtain a strict agreement for environmental values for two arcs that have difference of 10 years, approximation should be made.
Daily averaged $K_p$ and $F_{10.7}$ are considered for GOCE arc.
Arc for Q-Sat with comparable values is needed to be found.
The difference for the daily averaged value for $K_p$ and $F_{10.7}$ should be no larger than $0.6$ and $15$ respectively.
In Fig. \ref{EPs}, 3-hourly averaged $K_p$ and  24-hour averaged $F{10.7}$ from September 6th to September 21st in 2010 are plotted.
Q-sat arc on September 7, 10, 15, 17, October 2, November 1 in 2020 and February 26 on 2021 are selected and optimized Jacchia-Roberts by these arcs are used for GOCE satellite drag coefficient recognition.
Detailed mapping for which Q-Sat arc is employed for GOCE drag coefficient identification can be found in Tab.\ \ref{envMatch} and differences of daily averaged $K_p$ and $F_{10.7}$ between GOCE arc and Q-Sat arc are presented in Fig.\ \ref{EP_difference} in which $\Delta k_p = k_{p}(GOCE)-k_{p}(Q-Sat)$ and $\Delta F_{10.7} = F_{10.7}(GOCE)-F_{10.7}(Q-Sat)$.
\begin{table}[hhh!]
\caption{Mapping of the date. \label{envMatch}}
\begin{tabular}{@{}cc@{}}
\hline
GOCE Arc Date  & Q-Sat Arc Date \\
\hline
2010-09-06 & 2020-10-02 \\
2010-09-07 & 2020-10-02 \\
2010-09-08 & 2020-10-02 \\
2010-09-09 & 2020-10-02 \\
2010-09-10 & 2020-09-10 \\
2010-09-11 & 2020-09-10 \\
2010-09-12 & 2020-09-17 \\
2010-09-13 & 2020-09-07 \\
2010-09-14 & 2021-02-26 \\
2010-09-15 & 2020-09-15 \\
2010-09-16 & 2020-11-01 \\
2010-09-17 & 2020-10-02 \\ 
2010-09-18 & 2020-09-17 \\
2010-09-19 & 2020-09-17 \\
2010-09-20 & 2020-09-17 \\
\hline
\end{tabular}
\end{table}
Daily averaged $k_p$ is considered in here for relaxed matches.
Solar activity increased and decreased periodically with a 11-year cycle.
It can be observed that the the solar radiation 10 year ago is overall stronger than in year of 2020.

In present paper, as data from satellite that is currently orbiting is hard to reach, GOCE satellite is employed for purpose of demonstration, the orbit prediction procedure is complexified.
In real orbit prediction task using the proposed method, environmental parameters matching shown in this section is not required.

\FloatBarrier

\subsubsection{Orbit Data Length for Drag Coefficient Identification}
\begin{figure}[hhh!]
\centerline{
\includegraphics[width=3.5in]{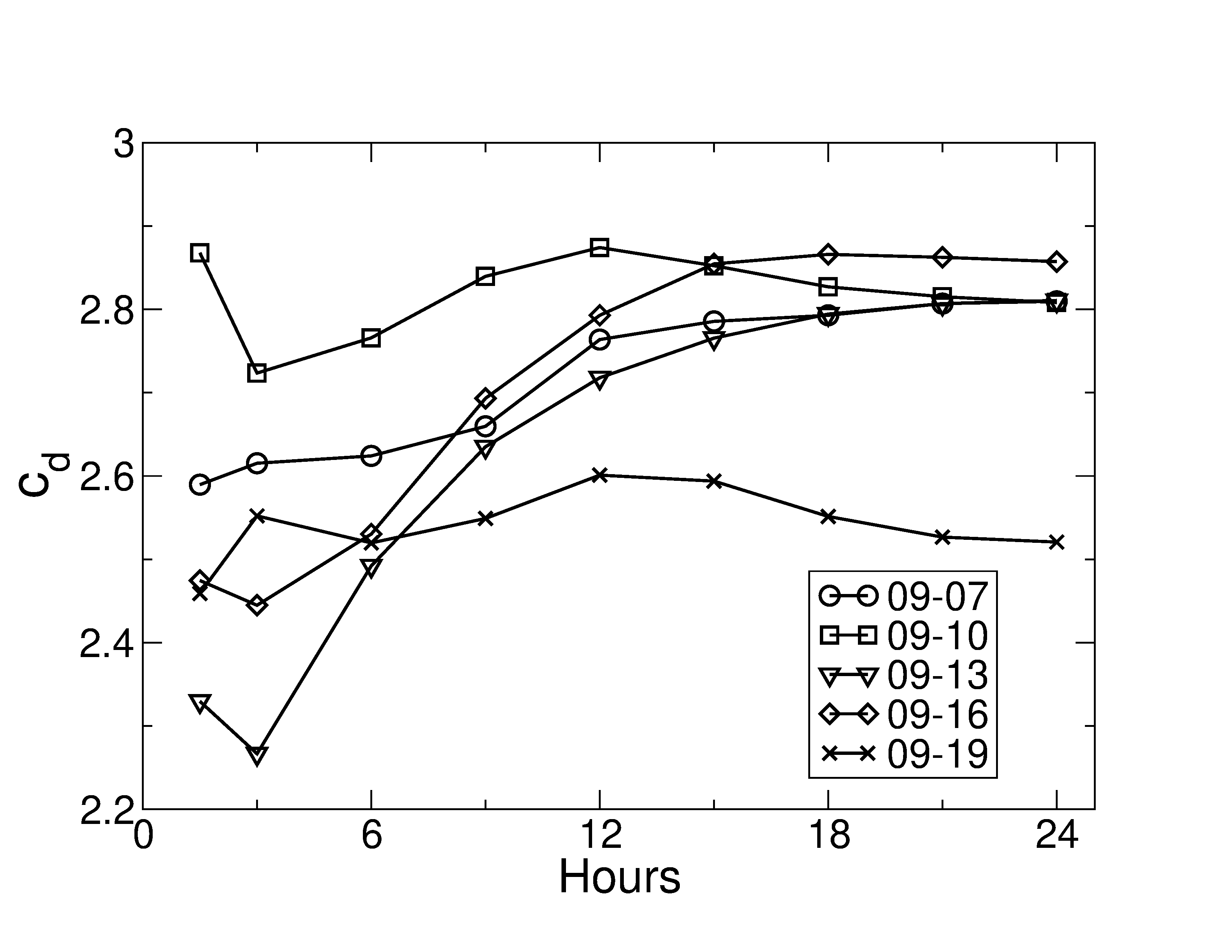}\noindent
}
\caption{
Drag coefficient calculated by different data length length.
}
\label{Drag}
\end{figure}

The dynamic approach-based inversion is employed for drag coefficient identification and the input for the inversion is the orbit data of the spacecraft which in needs of trajectory prediction.
Optimized EADMs by corresponding Q-Sat orbit data (Tab. \ref{envMatch}) is implemented in the inversion.
To save computational expense and produce a rapid prediction, arclength of the last $1.5$, $3$, $6$, $9$, $12$, $15$, $18$, $21$, $24$ hours of the first day are tested for finding an optimal trajectory length for drag coefficient identification.
In Fig.\ \ref{Drag}, drag coefficient calculated by data on September 07, 10, 13, 16, 19 are plotted.
For all cases, the drag coefficient approached to a certain value with increased data and converged to values calculated by 24-hour data.
For example, with 21-hour orbit data on September 7th, drag coefficient is coming close to 2.8.
For data on September 16, with 18-hour data, drag coefficient is approximately 2.86 which is around the value generated by 24-hour data.
As a consequence, drag coefficient generated by arc with length of 24 hours is chosen for the orbit prediction performed in later section. 
The results obtained here is comparable to the result calculated by direct simulation Monte Carlo method by Koppenwallner \cite{koppenwallner_satellite_2011}.
Orbit environmental parameters such as temperature, species of the particles it encountered, etc., can alter the value of the drag coefficient \cite{bernstein_evidence_2020,moe_gassurface_2005}.
The drag coefficient obtained here is an averaged value over a certain amount of time and it contains all the uncertainties occur during the period.
It can represent drag coefficient of the next day since environmental perturbations of the next day often show a good match with those of the current day \cite{jaggi_goce_2011}.

\subsection{Spacecraft Orbit Prediction}
\begin{figure}[hhh!]
\centerline{
\includegraphics[width=3.5in]{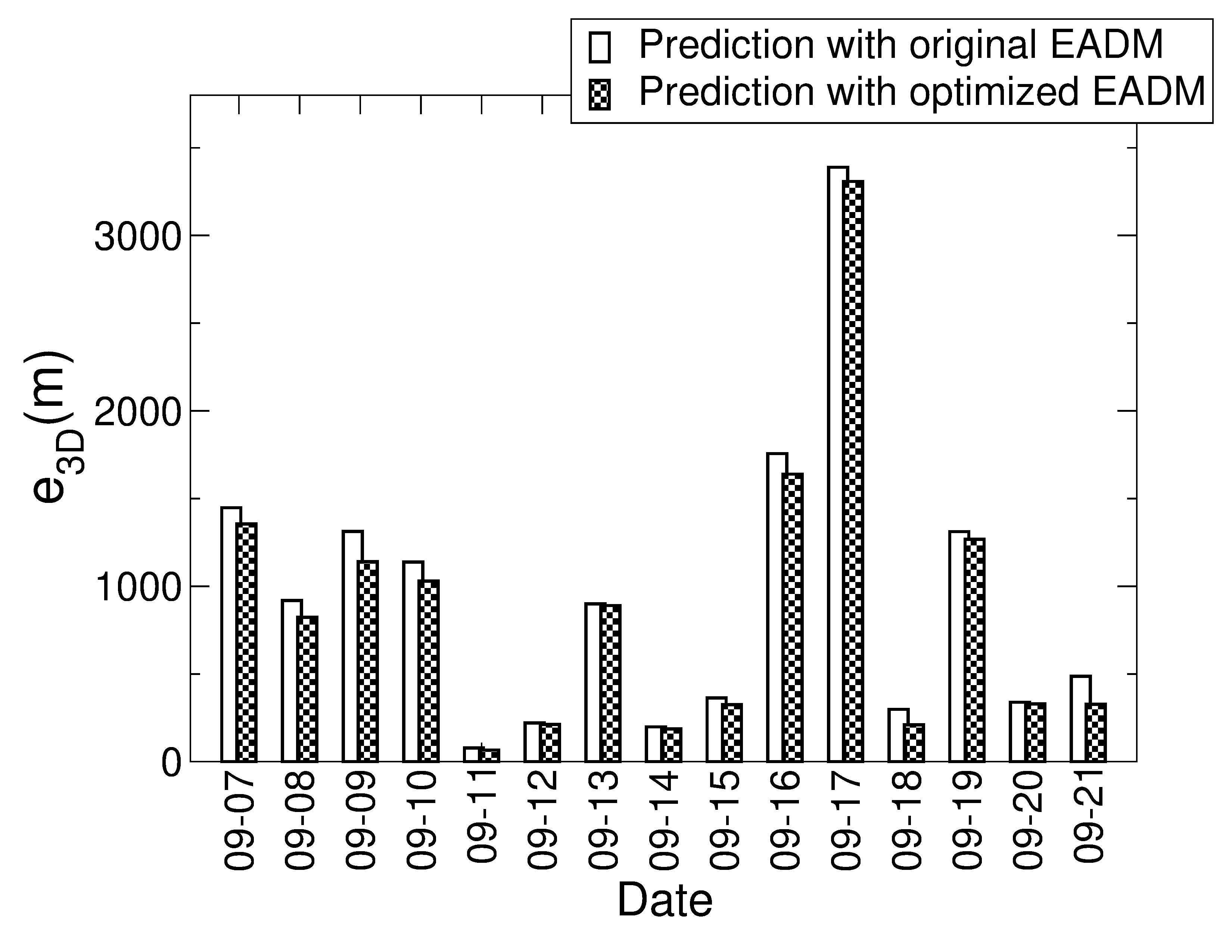}\noindent
}
\caption{
Maximum 3D error for prediction interval of 24 hours.
}
\label{predictionError}
\end{figure}

\begin{figure}[hhh!]
\centerline{
\includegraphics[width=3.5in]{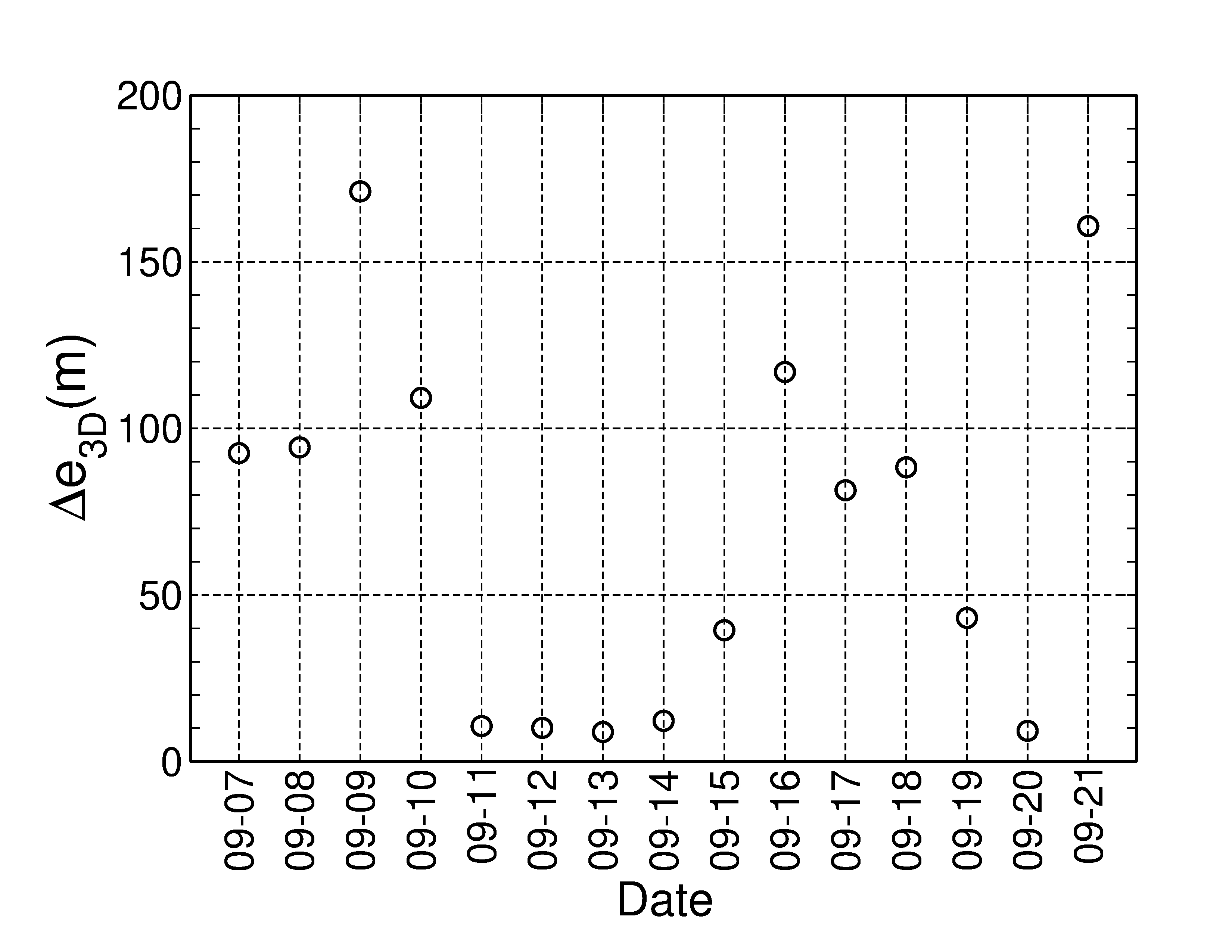}\noindent
}
\caption{
Difference of the maximum 3D error between prediction with original EADM and with optimized EDAM.
}
\label{compare}
\end{figure}

With the identified drag coefficient and optimized EADM, orbit predictions are conducted by extrapolation with the dynamic parameters employed for drag coefficient identification.
The prediction extend for 24 hours and the predicted orbit is compared with the actual orbit.
The 3D error,
\begin{equation}
e_{3D}=\sqrt{\sum_{i=1}^3 ( x_{i,p} - x_{i,a}) ^2 },
\end{equation}
is calculated where $i=1,2,3$ which represents the along-track, cross-track and radial direction and subscript '$p$' represents for predicted orbit and '$a$' for actual orbit.
Maximum 3D prediction errors in the 24-hour interval are found and plotted in Fig.\ \ref{predictionError}. 
For comparison, maximum 3D error for orbit prediction with original Jacchia-Roberts model are also included.
As can be observed in Fig. \ref{predictionError}, prediction with optimized EADM has overall decreased maximum 3D errors.
It indicated an overall higher predication accuracy.
The difference between the two prediction, $\Delta e_{3D}$, are plotted in Fig. \ref{compare}.
The GOCE satellite 24-hour prediction accuracy is increased by 171m at the highest.
The averaged 24-hour prediction accuracy over 14 days is increased approximately 70m.
It can be concluded that the accuracy of the orbit prediction for GOCE satellites was improved significantly.

\FloatBarrier

\section{Conclusion}

To diminish the convoluted uncertainties in the correction prediction strategy which based on single spacecraft orbital behavior, a Q-Sat trajectory based LEO spacecraft orbit prediction method is proposed in present paper.
The Q-Sat satellite, launched by xxxxx lab at xxx University which initially orbit around 500km, is a spherical micro satellite for gravity field recovery and atmosphere density detection.
The satellite is considered to have a constant drag coefficient of $2.2$ regardless its attitude and its tracking data is employed for EADM modification by dynamic approach-based inversion introduced in present paper.
The revised EADM is later employed in drag coefficient identification process for the spacecraft which orbit needs to be predicted.
Convoluted error is thus reduced as EDAM modification and drag coefficient recognition are decoupled.
As can be seen in many research that, in actual space, spacecraft drag coefficient changes due to variations such as composition of gas, gas-surface interaction, molecules absorption, etc. 
In current work, drag coefficient of the GOCE satellite is calculated in a daily averaged fashion, which contains the uncertainties occured in an entire day.
As a result, the identified drag coefficient of the GOCE satellite is no longer identical from a day to another.
So as to revised EADM that varied from day to day.
The obtained drag coefficient and EADM with space environment parameters from the previous day are then applied in the trajectory forecast for next 24 hours as the space environment can be considered similar as of the previous day.
For demonstration of the proposed method, space environment parameters $k_p$ and $F_{10.7}$ for arc of GOCE and Q-Sat should be comparable in order to obtain a higher prediction acurracy.
In real prediction workflow, this step in not necessary.
However, forecast of the space environment parameters for the next day is of also important for improving prediction accuracy.
Otherwise, values from previous day will need to be employed.
In present work, the GOCE satellite 24-hour prediction accuracy is increased by 171m at the highest compared to result by legacy "correction-prediction" strategy.
The 14 days averaged 24-hour prediction accuracy is increased approximately 70m. 

\section*{Data Availability}
The datasets employed during the current study are available from the corresponding author on reasonable request.

\section*{Code availability}
The custom codes for current research are available from the corresponding author upon request.

\printbibliography

\section*{Competing interests}
The authors declare no competing interests.

\setcounter{figure}{0}

\end{document}